\documentclass{emulateapj}
\usepackage{apjfonts,times}
\newcommand{\beq}{\begin{equation}}
\newcommand{\eeq}{\end{equation}}
\def\farcm{\hbox{$.\mkern-4mu^\prime$}}
\def\farcs{\hbox{$.\!\!^{\prime\prime}$}}
\def\earth{\hbox{$\oplus$}}
\def\arcmin{\hbox{$^\prime$}}
\def\arcsec{\hbox{$^{\prime\prime}$}}

\newcommand{\AmS}{{\protect\the\textfont2
  A\kern-.1667em\lower.5ex\hbox{M}\kern-.125emS}}
\newcommand{\lsim}{\ \raise
-2.truept\hbox{\rlap{\hbox{$\sim$}}\raise5.truept\hbox{$<$}\ }}
\newcommand{\gsim}{\ \rais-2.truept\hbox{\rlap{\hbox{$\sim$}}\raise5.truept\hbox{$>$}\ }}
\newcommand{\simsim}{\ \raise
-2.truept\hbox{\rlap{\hbox{$\sim$}}\raise5.truept\hbox{$\sim$}\ }}

\slugcomment{Accepted for Publication in ApJ --- Draft Version April 
22, 2007}

\shorttitle{HST/ACS observations of the intermediate-age cluster BS 90
in the SMC}
\shortauthors{Rochau B., et al.}

\begin{document}

\title{The Star-forming Region NGC 346 in the Small Magellanic Cloud with 
Hubble Space Telescope ACS Observations. II. Photometric Study of the 
Intermediate-Age Star Cluster BS 90\footnote{Research supported by the 
Deutsche Forschungsgemeinschaft (German Research Foundation)}
}

%\footnotetext[]{Research supported by the Deutsche 
%Forschungsgemeinschaft (German Research Foundation)}
%\altaffiltext{1}{Research supported by the Deutsche 
%Forschungsgemeinschaft (German Research Foundation)}

\author{Boyke Rochau and Dimitrios A. Gouliermis}
\affil{Max Planck Institute for Astronomy, K\"onigstuhl 17, 69117
Heidelberg, Germany}
\email{rochau@mpia.de, dgoulier@mpia.de}

\author{Wolfgang Brandner}
\affil{UCLA, Div. of Astronomy, 475 Portola Plaza, Los Angeles, CA
90095-1547, USA}
\affil{Max Planck Institute for Astronomy, K\"onigstuhl 17, 69117
Heidelberg, Germany}
\email{brandner@astro.ucla.edu, brandner@mpia.de}

\author{Andrew E. Dolphin}
\affil{Steward Observatory, University of Arizona, Tucson, AZ 85721, USA}
\affil{Raytheon Corporation, USA}
\email{adolphin@raytheon.com}

\and

\author{Thomas Henning}
\affil{Max Planck Institute for Astronomy, K\"onigstuhl 17, 69117
Heidelberg, Germany}
\email{henning@mpia.de}

\begin{abstract}

We present the results of our investigation of the intermediate-age star 
cluster BS~90, located in the vicinity of the {\sc H ii} region N 66 in 
the SMC, observed with HST/ACS. The high-resolution data provide a unique 
opportunity for a very detailed photometric study performed on one of the 
rare intermediate-age rich SMC clusters. The complete set of observations 
is centered on the association NGC 346 and contains almost 100,000 stars 
down to $V \simeq\ 28$ mag. In this study we focus on the northern part of 
the region, which covers almost the whole stellar content of BS 90. We 
construct its stellar surface density profile and derive structural 
parameters. Isochrone fits on the CMD of the cluster results in an age of 
about $4.5$ Gyr. The luminosity function is constructed and the 
present-day mass function of BS 90 has been obtained using the 
mass-luminosity relation, derived from the isochrone models. We found a 
slope between $-1.30$ and $-0.95$, comparable or somewhat shallower than a 
typical Salpeter IMF. Examination of the radial dependence of the mass 
function shows a steeper slope at larger radial distances, indicating mass 
segregation in the cluster. The derived half-mass relaxation time of 
$0.95$ Gyr suggests that the cluster is mass segregated due to its 
dynamical evolution. From the isochrone model fits we derive a metallicity 
for BS 90 of $\rm{[Fe/H]}=-0.72$, which adds an important point to the 
age-metallicity relation of the SMC. We discuss our findings on this 
relation in comparison to other SMC clusters.

\end{abstract}


\section{Introduction}\label{intro}

Studies of star clusters, covering a wide range of ages, metallicities
and environments offer the opportunity to investigate the evolution of
individual clusters, but also of the entire parent galaxy, e.g. its star
formation history and chemical enrichment. The clusters of the Large
and Small Magellanic Cloud (LMC, SMC) are excellent targets for such studies.
Their proximity allows to resolve individual members of a stellar system
and to assume that all stars of a star cluster are more or less located
at the same distance. Studies of cluster systems of both the LMC and SMC
provide the opportunity to test theories of stellar evolution for 
different ages, abundances and/or mass content in an environment
different than the Milky Way.  Star clusters of the Magellanic Clouds
(MCs) have ages varying from the early stages of the clouds up to the
present, therefore allowing the study of the evolution of the clouds
from the time they have been formed until today. 

The history of cluster formation in the SMC is the topic
of ongoing discussion. Although the cluster population has been
considered as having a continuous age distribution (Da Costa \&
Hatzidimitriou 1998; Mighell et al. 1998), some studies suggest that most
of the SMC clusters have been formed in two epochs. Rich et al. (2000)
found seven populous SMC star clusters being formed either around $8\pm
2$ Gyr or around $2\pm 0.5$ Gyr ago. Rafelski \& Zaritsky (2005) further
argue that the cluster age distribution shows few peaks, but no evidence
for a significant age gap in the SMC as observed in the LMC (van den Bergh 1991; 
Da Costa 1991; Westerlund 1997). They suggest that the dominant
initial epoch of cluster formation together with a fast dissolution of
clusters makes the age distribution of the SMC clusters to appear with a
quiescent cluster formation period in the intermediate-age range.
Consequently, the intermediate-age range in the SMC is sparsely
populated.

The low number of intermediate-age SMC clusters makes it difficult to 
establish a consistent model for its chemical enrichment history. Pagel \& 
Tautvai\v{s}ien\'{e} (1998) published two models to explain the enrichment 
of the cluster population of both the SMC and LMC. Their work includes a 
`bursting' model of star formation and a model with `smooth' star 
formation. For the latter, a constant star formation rate is assumed over 
the entire lifetime of the clouds. It is comparable to a model, which has 
been previously published by Da Costa \& Hatzidimitriou (1998). The 
bursting model assumes a constant star formation rate over a certain time 
interval with a discontinuous change between two bursts. Pagel \& 
Tautvai\v{s}ien\'{e} (1998) included two bursts for the SMC, one at the 
time of its formation and the second at $\sim 2.5$ Gyr ago. Ages and 
metallicities of the cluster population have been examined with 
photometric analyses (e.g. Mighell et al. 1998; Piatti et al. 2001, 2005a) 
or via spectroscopy (e.g.  Da Costa \& Hatzidimitriou 1998; de Freitas 
Pacheco et al. 1998; Piatti et al. 2005b). Results of this studies are not 
providing a clear picture of the chemical enrichment of the SMC.  As an 
example, results from Mighell et al. (1998) are in better agreement with 
the burst model of Pagel \& Tautvai\v{s}ien\'{e} (1998), whereas Da Costa 
\& Hatzidimitriou (1998) pointed out that their findings are consistent 
with a more continuous chemical evolution following a simple `closed-box' 
model for the SMC, but they also mentioned that it may be an `open' rather 
than a closed box due to its interaction with the LMC and the Galaxy. 
Naturally, the further investigation of intermediate-age star clusters 
would help to clarify the history of the cluster population of the SMC.

Recent high-resolution imaging with the Advanced Camera for Surveys (ACS) 
on-board the Hubble Space Telescope (HST) of the general area of the 
stellar association NGC 346 in the SMC includes the intermediate-age star 
cluster BS[95]~90 (Bica \& Schmitt 1995) or in short BS~90, providing, 
thus, a unique opportunity to study one of the very rare star clusters, 
which has been formed in a possible quiescent epoch of star formation in 
the SMC. Such high-resolution studies are quite rare.  The catalog of SMC 
clusters imaged with the Wide-Field Planetary Camera 2 on-board HST, 
presented by Mackey \& Gilmore (2003b), includes only 10 clusters, seven 
of which have ages between 12 Gyr and 1.4 Gyr. The study we present here 
on the cluster BS 90 is based on a unique set of deep observations toward 
a SMC cluster with the high resolving power of HST/ACS. BS 90 is located 
in the vicinity of the stellar association NGC 346, the largest stellar 
concentration in the SMC, which is related to LHA 115-N 66 (in short N 
66), the brightest {\sc H ii} region of the galaxy (Henize 1956). NGC 346 
has been subject of several studies. Peimbert et al. (2000) studied the 
chemical composition of the region. High-sensitivity CO and H$_{\rm 2}$ 
observations and imaging in [{\sc O iii}] of the photodissociation region 
is presented by Rubio et al. (2000). The X-ray emission in the general 
area has been studied by Naz{\'e} et al. (2003, 2004). Finally, a snapshot 
of the star formation history of the entire region is recently presented 
by Sabbi et al. (2007).

In the first paper of our investigation of the region of NGC 346 with 
HST/ACS observations (Gouliermis et al. 2006, from here-on Paper I) we 
present our photometry and the spatial distribution of different stellar 
populations in the whole region, revealing the stellar richness of the 
cluster BS 90. Our photometry allowed to resolve stars down to $V\simeq28$ 
mag ($\approx5$ mag below the turn-off), corresponding to $\sim0.4\ 
\rm{M_{\odot}}$. Naturally, the inclusion of BS 90 in this data set offers 
the opportunity for a detailed analysis of this unique cluster using one 
of the richest stellar samples in the SMC. Here, we present our results of 
such an analysis. The structure of the paper is the following. In 
\S~\ref{data} we discuss the data set and the data reduction. The 
construction and analysis of the stellar surface density of the cluster is 
presented in \S~\ref{structure}. In \S~\ref{cmdsection} the 
color-magnitude diagram (CMD) of BS 90 is constructed and its properties 
are derived by isochrone fitting.  In \S~\ref{lfmf} the luminosity and 
mass functions of the cluster are constructed and presented, and we 
investigate the radial dependence of the mass function to examine a 
possible mass segregation in \S~\ref{segregation}. We discuss our findings 
concerning the age and metallicity of BS 90 in terms of the chemical 
enrichment history of SMC in \S~\ref{ceh-smc}. Finally, in \S~\ref{sum} we 
summarize our investigation of this unique SMC cluster.

%\clearpage
%%%%%%%%%%%%%%%%%%%%%%%%%%%% FIGURE %%%%%%%%%%%%%%%%%%%%%%%%%%%%%%%%%%%%%%%
\begin{figure}[t!]
\epsscale{1.25}
\plotone{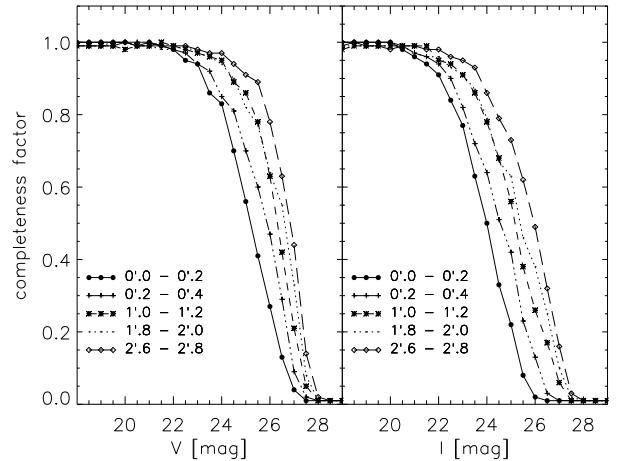}
\caption{Completeness functions in $V$ ({\em left}) and $I$ ({\em
right}) for five different distances from the center of the cluster.
Annuli with radii of $0\farcm0-0\farcm2$, $0\farcm2-0\farcm4$,
$1\farcm0-1\farcm2$, $1\farcm8-2\farcm0$, $2\farcm6-2\farcm8$ have been
selected.}
\label{fig_compl}
\end{figure}
%%%%%%%%%%%%%%%%%%%%%%%%%%%%%%%%%%%%%%%%%%%%%%%%%%%%%%%%%%%%%%%%%%%%%%%%%%%

\section{Data Description and Analysis}\label{data}

The data we use in this study are collected within the HST Program
GO-10248. Three pointings were observed using the Wide-Field Channel
(WFC) of ACS, centered on the association NGC 346, in the broad-band
filters $F555W$ and $F814W$, equivalent to standard $V$ and $I$
respectively. The data sets are retrieved from the HST Data Archive.
Photometry of the pipeline-reduced FITS files was performed using the
ACS module of the package {\tt DOLPHOT} (version 1.0). This mode of the
package is an adaption of {\tt HSTphot} (Dolphin 2000), especially
designed for photometry on ACS imaging. With our photometry we detected
almost 100,000 stars. A detailed description of the datasets and a full
account of the photometric process is given in Paper I, where we also
made the whole photometric stellar catalog available.

This catalog lists the celestial coordinates, and $V$- and
$I$-magnitudes (in the Vega system) with the corresponding photometric
errors for each star.  Typical photometric uncertainties as a function
of the magnitude are shown in Paper I (Figure 2). The completeness of
the data has been calculated by running {\tt DOLPHOT} in artificial-star
mode as described in Paper I, and artificial star lists were created
with the utility {\em acsfakelist}. Completeness depends on the level of
crowding, and, consequently, in a populous compact cluster like BS 90 it
should vary with respect to the distance from the center of the cluster,
where crowding is higher. Indeed, the completeness in the area of BS 90
is found to be spatially variable. This is seen in Fig.~\ref{fig_compl},
where the completeness function in $V$ and $I$ (with the completeness
factors estimated as a function of magnitude as $N_{\rm detected}/N_{\rm
simulated}$) is shown for different selected distances from the center
of the cluster.

\section{Dynamical Behavior of the Cluster}\label{structure}

\subsection{Stellar Surface Density Profile of BS 90}\label{ssdp}

The observed field covers a quite large area around BS 90, allowing the
construction of the stellar surface density profile of the cluster at relatively 
large distances from its center. We divided the area of the cluster in
30 concentric annuli with steps of $6\arcsec$ and we counted the number of stars
within each annulus. We corrected the counted stellar numbers for
incompleteness according to the corresponding completeness factors.
Completeness is a function of both distance from the center of the
cluster and magnitude.  If $N_{i}(V)$ is the number of detected stars in
the $V$-band within the $i^{th}$ annulus and in the brightness range
corresponding to the completeness factor $c_{i}(V)$, the completeness
corrected number of stars is \begin{equation} N_{i,c}=\sum_{V}^{}
c_{i}(V)\cdot N_{i}(V). \end{equation} We obtained the stellar surface
density, $f_{i}$, by normalizing this number to the area of the
corresponding annulus as $f_{i}=N_{i,c}/A_{i}$, with $A_{i}$ being the
area of the $i^{th}$ annulus. The observed field-of-view does not fully
cover the northern and western extend of the cluster, and, therefore, for
the annuli not completely observed, we considered only the available area
for the estimation of the corresponding surface.

\begin{figure}[t!]
\epsscale{1.2}
\plotone{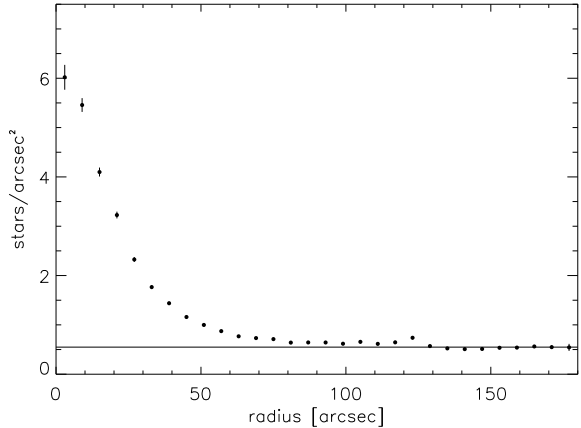}
\caption{Radial stellar surface density profile. The stellar numbers
have been corrected for incompleteness. The solid line shows the uniform
level of the stellar background density, which corresponds to a value of
$\sim0.55~{\rm stars}/{\Box \arcsec}$. The errors represent the
uncertainties due to Poisson statistics. The small density peak at
distance $\sim 120\arcsec$ is due to a young compact cluster to the east
of BS 90.}
\label{fig-dpl}
\end{figure}

The constructed stellar surface density profile is shown in
Fig.~\ref{fig-dpl}.  The errors correspond to uncertainties due to the
counting process and they represent the Poisson statistics. As expected,
this density profile shows a smooth drop as a function of distance from
the center of the cluster. It drops to a uniform level, with the
exception of the small increase at the distance around $2\arcmin$. We
identify this increase as a small young cluster located east of BS 90.
The horizontal uniform level represents the stellar density of the field
of the galaxy. Its value is shown as the solid line in
Fig.~\ref{fig-dpl} and has been measured by fitting the models of Elson,
Fall \& Freeman (1987) to the stellar surface density profile as shown
below.

%\clearpage
%%%%%%%%%%%%%%%%%%%%%%%%%%%% FIGURE %%%%%%%%%%%%%%%%%%%%%%%%%%%%%%%%%%%%%%%
%%%%%%%%%%%%%%%%%%%%%%%%%%%%%%%%%%%%%%%%%%%%%%%%%%%%%%%%%%%%%%%%%%%%%%%%%%%

\subsection{Structural Parameters}\label{parameters}

We applied both the empirical model by King (1962) and the model of
Elson, Fall \& Freeman (1987; from here-on EFF) to the stellar surface
density profile of BS 90 in order to obtain its structural parameters.
The EFF model is suited for clusters without tidal truncation, while
King's empirical model represents a tidally truncated cluster.  Both
models provide the opportunity to derive accurate characteristic radii
for the cluster, such as its core and tidal radius. The core radius ($r_{\rm
c}$) describes the distance from the center of the cluster where the
stellar density drops to the half of its central value and the tidal radius 
($r_{\rm t}$) is the limit where the density drops to zero. For the
application of King's model the stellar surface density of the cluster
alone (with no contamination from the field) is necessary, while the EFF
model does not require any field subtraction. We first apply the latter
in order to estimate the core radius and the uniform background level. 
The subtraction of this density level from the measured surface density
at each annulus gives the surface density profile of the cluster alone,
from which we derive the core and tidal radii.

\subsubsection{Best Fitting EFF Profile}

\begin{figure*}[t!]
\epsscale{1.}
\plotone{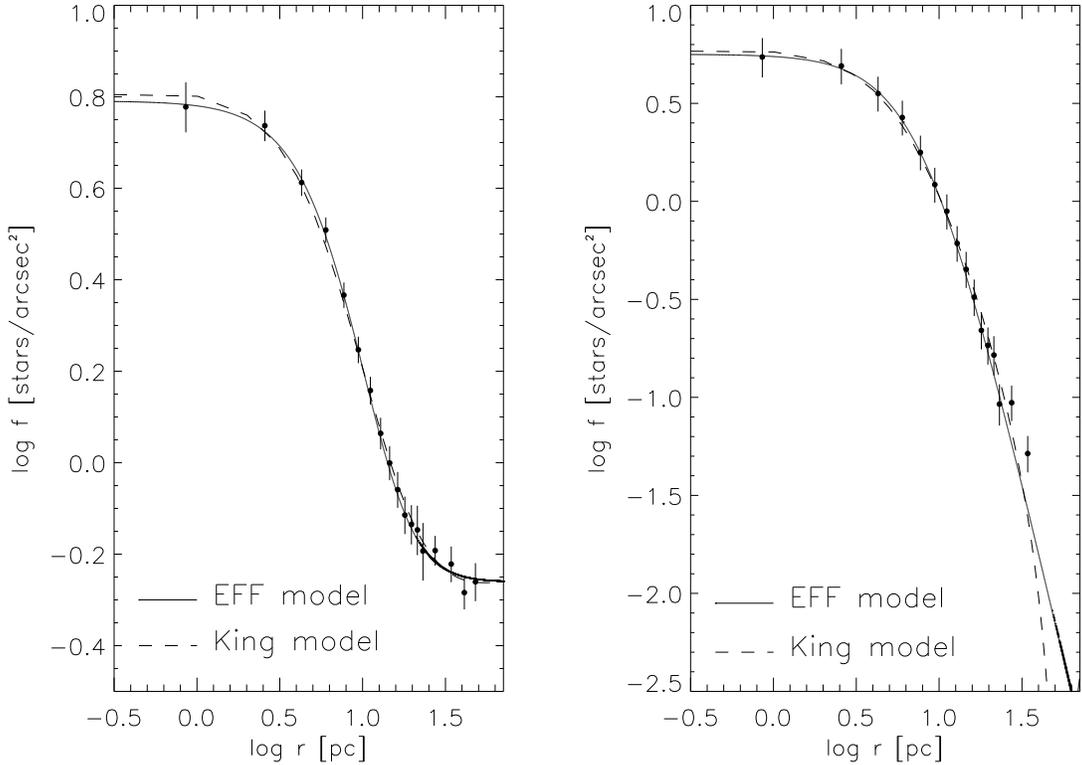}
\caption{Radial surface density profile of BS 90 with the best fitting 
King and EFF models superimposed. The King model is overplotted with a 
dashed line and the EFF model with a solid line. The error bars represent 
Poisson statistics. {\em Left}: The profile with all detected stars, 
including field and association. {\em Right}: The profile of the cluster 
alone after subtracting the estimated background density level. Both 
models fit very well the density profiles with a core radius of $r_{\rm 
c}\simeq5.6$ pc (EFF) or $r_{\rm c}\simeq5.4$ pc (King), a power-law slope 
of $\gamma=3.68$ (EFF), and a tidal radius of $r_{\rm t}\simeq53.7$ pc 
(King).}
\label{fig-king}
\end{figure*}

From studies of young LMC clusters by EFF it appears that the examined
clusters are not tidally truncated. These authors developed a model more
suitable to describe the stellar surface density profile of such
clusters:
\begin{equation}
f(r)=f_{0}(1+r^{2}/a^{2})^{-\gamma/2}+f_{\rm{field}}
\label{eq-elson}
\end{equation}
where $f_{0}$ is the central stellar surface density, $a$ is a measure
of the core radius and $\gamma$ is the power-law slope which describes
the decrease of surface density of the cluster at large radii
($f(r)\propto r^{-\gamma/2}$ for $r\gg a$).  $f_{\rm{field}}$ represents
the uniform background density level. Our best fit of the EFF model to
the surface density profile of BS 90 gives $a=8.24\pm0.67$ pc and
$\gamma=3.68\pm1.05$. The density of the background field, taken from the
fitting procedure, corresponds to a value of
$f_{\rm{field}}\sim0.55\pm0.04~{\rm stars}/{\Box \arcsec}$. According to
EFF model, the core radius, $r_{\rm c}$, is given from Eq.~\ref{eq-elson}
assuming no contribution from the field as:
\begin{equation}
r_{\rm c}=a(2^{2/\gamma}-1)^{1/2} \label{eq-rc} 
\end{equation} 
We estimate a core radius of $r_c=5.57^{+0.73}_{-0.37}$ pc. The
best-fitting EFF model assuming the derived values of $a, \gamma\
\rm{and}$ $f_{\rm{field}}$ is shown superimposed on the stellar surface
density profile in Fig.~\ref{fig-king}.

%\clearpage
%%%%%%%%%%%%%%%%%%%%%%%%%%%% FIGURE %%%%%%%%%%%%%%%%%%%%%%%%%%%%%%%%%%%%%%%
%%%%%%%%%%%%%%%%%%%%%%%%%%%%%%%%%%%%%%%%%%%%%%%%%%%%%%%%%%%%%%%%%%%%%%%%%%%

\subsubsection{Best fitting King Profile}\label{bestfitking}

In order to have the density profile of the cluster alone we subtracted
the background density level of $\sim0.55\pm0.04~{\rm stars}/{\Box
\arcsec}$, estimated above, from the profile of Fig.~\ref{fig-dpl}. Then
we use the field-subtracted profile to derive the core and tidal radius
of the cluster as described by King (1962).  Specifically according to this 
model the core radius is found from the formula
\begin{equation}
f=\frac{f_0}{1+(r/r_{\rm c})^2}
\label{core}
\end{equation}
which describes the inner region of the cluster and the tidal radius from
\begin{equation}
f=f_1(1/r-1/r_{\rm t})^2
\label{tidal}
\end{equation}
which represents the outskirts of the cluster.
$f_0$ describes again the central surface density of the cluster and
$f_1$ is a constant. The best fitting profile results in a tidal radius
$r_{\rm t}=53.66^{+8.93}_{-7.10}$ pc
($3\farcm13^{+0\farcm53}_{-0\farcm41}$) and a core radius $r_{\rm
c}=5.44^{+0.33}_{-0.35}$ pc ($19\farcs06^{+1\farcs15}_{-1\farcs23}$).
The concentration parameter, $c$, is defined as the logarithmic ratio of
tidal to core radius $c=\log{(r_{\rm t}/r_{\rm c})}$ and refers to the
compactness of the cluster.  We obtained a concentration parameter
$c=1.0\pm0.1$. According to King's model the density profile of a
tidally truncated cluster is given as:
\begin{equation}
f\propto\Bigl(\frac{1}{[1+(\frac{r}{r_{\rm c}})^2]^\frac{1}{2}}-\frac{1}
{[1+(\frac{r_{\rm t}}{r_{\rm c}})^2]^\frac{1}{2}}\Bigr)^{2}
\label{eq-king}
\end{equation}
where $f$ is the stellar surface density, $r_{\rm c}$ and $r_{\rm t}$
the core and tidal radius, respectively and $r$ is the distance from the
center. The best fitting King model is shown as the dashed line,
superimposed on the radial surface density profile in
Fig.~\ref{fig-king}.

Both King and EFF models are in good agreement to each other.
The intermediate region is equally well fitted by both models. Based on the
accuracy of the fits, it can be concluded that the EFF model is in better agreement
with our data in the inner part of the cluster but for the intermediate and 
outer regions of the cluster the best fits yield no preference 
for the tidally truncated King model or the EFF model. 

\subsubsection{Core Radius-Age Relation of SMC clusters}

\begin{figure}[t!]
\epsscale{1.25}
\plotone{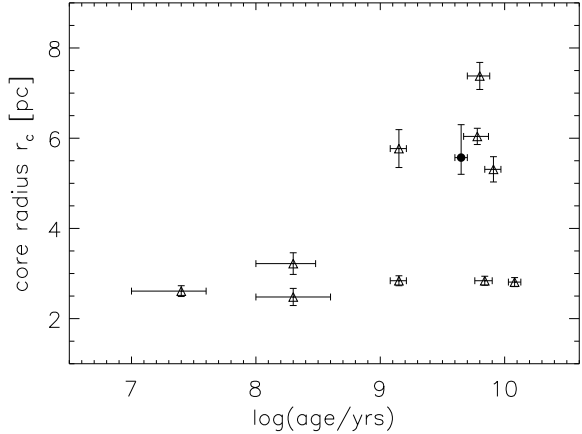}
\caption{Core radius versus age for the SMC clusters examined by Mackey
\& Gilmore (2003b) with the values for BS 90 found in this paper also
plotted. The triangles represent the clusters which have been
investigated by Mackey \& Gilmore (2003b). The star cluster BS 90 is   
displayed as the black dot. Clusters which are located in the upper
right may have not been evolved in isolation.}
\label{fig-core_age}
\end{figure}

A comparison of 10 SMC clusters, investigated by Mackey \& Gilmore
(2003b), shows that they can be divided into two groups with respect to
their core radii, as they are derived from the EFF model. One group
contains six star clusters (NGC 176, 330, 121, 411, 416, 458) with radii
less than $5$ pc and the second group includes the remaining four clusters (NGC
361, 152, 339, Kron 3) with $r_{\rm c}>5$ pc.  BS 90 with its core
radius of $r_{\rm c}=5.57$ pc, as found from EFF models, would be a
member of the second group. Mackey \& Gilmore examined the relationship
between the core radii and the ages of clusters located in the LMC, SMC,
Fornax and Sagittarius dwarf spheroidals (see Mackey \& Gilmore
2003a,b,c), and in Fig.~\ref{fig-core_age} the core radius-age
relationship for the SMC cluster sample is shown with our measurements
for BS 90. The distinction of the clusters into the two groups according
to their core radii is easily seen in this plot. For the LMC and the
Fornax and Sagittarius dwarf spheroidals they observed a similar
distinction and argued that it shows a real evolution of cluster
structure with age (Mackey \& Gilmore 2003a,c), with the sequence of
clusters with smaller core radii following the `standard' isolated
globular cluster evolution. These authors suggest that the evolution of
clusters at the upper right of Fig.~\ref{fig-core_age}, where BS 90
belongs to, differ from this isolated evolution. The disparity could be due
to very different stellar populations in these clusters or because they
do not follow an isolated evolution, but rather an evolution influenced
by external processes.

%\clearpage   
%%%%%%%%%%%%%%%%%%%%%%%%%%%% FIGURE %%%%%%%%%%%%%%%%%%%%%%%%%%%%%%%%%%%%%%%
%%%%%%%%%%%%%%%%%%%%%%%%%%%%%%%%%%%%%%%%%%%%%%%%%%%%%%%%%%%%%%%%%%%%%%%%%%%

McLaughlin \& van der Marel (2005) also derived core and tidal radii
with the application of King empirical models and the same separation of
the clusters into the two groups according to their core radii was
observed in their sample. Our application of King empirical models
yields a core radius of $r_{\rm c}\simeq5.4$ pc and a tidal radius of
$r_{\rm t}\simeq53.7$ pc, making BS 90 also a member of to the second group of the
McLaughlin \& van der Marel sample of clusters. Our estimated
concentration parameter for BS 90 $c\simeq1.0$ is relatively low in comparison to the SMC
clusters studied by Mackey \& Gilmore (2003b) and McLaughlin \& van der
Marel (2005). 

\subsubsection{Half-Light Radius}\label{halflight}

The half-light radius of the cluster is derived from its surface
brightness profile. It corresponds more or less to the half-mass radius and 
the corresponding half-mass relaxation time (\S~\ref{halfmass}) is a useful 
reference time for the dynamical evolution of the cluster as a whole 
(Spitzer 1987). To construct the surface brightness profile we subdivided the area of the
cluster into 26 annuli. The first 22 annuli were selected in steps of
$6\arcsec$ and the remaining four in steps of $12\arcsec$. We changed
the width of the annuli to improve the statistics in the outer regions
of the cluster. We converted the measured apparent magnitudes into flux
and counted the brightness of the stars within each annulus. The total
brightness of each annulus has been corrected for incompleteness and
normalized to the corresponding area. From our fit of the profile with
the EFF model we derived the total brightness of the entire cluster and
further obtained its half-light radius. We found a half-light radius of
$r_{\rm h}=9.42\pm0.5$ pc. 

\section{Color-Magnitude diagram}\label{cmdsection}

\begin{figure}[t!]
\epsscale{1.25}
\plotone{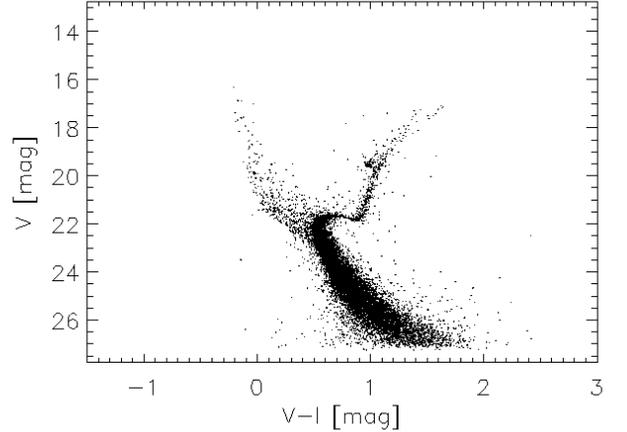}
\caption{$V$, $V-I$ CMD of all detected stars within a circular area of
$2\ r_{\rm c}$, corresponding to $\sim 11.2$ pc, around the center of
the cluster BS 90. The turn-off is clearly visible and the features of
the red clump, the sub-giant branch and the red giant branch are easily
seen as well, making BS 90 a {\em bona-fide} case. This CMD is
contaminated by field stars and also by the close-by association NGC 346,
the population of which is apparent in the upper main sequence with
colors bluer $V-I\lesssim0.5$ mag.}
\label{fig-cmd_all}
\end{figure}

The $V-I$ vs. $V$ color-magnitude diagram (CMD) of all stars detected
within a circular area of two core radii around the center of
the cluster is shown in Fig.~\ref{fig-cmd_all}.  A comparison of the
contour maps of the region around NGC 346, constructed with star counts
in Paper I, shows that the main body of the cluster covers well the
northern part of the observed field-of-view (Paper I, Figure 4). Our
selected radius of 2 $\times$ $r_{\rm c}$ around the center of BS 90
corresponds to $11.15$ pc and is comparable to the half-light radius of
$9.42$ pc.  It contains more than 40\% of the total number of observed stars and
more than 60\% of the observed cluster population. The CMD of
Fig.~\ref{fig-cmd_all} is indicative of an old cluster with a clear
turn-off at $V\sim22$ mag and $V-I\sim0.5$ mag, a well defined sub-giant
branch between $0.6 \lesssim V-I\lesssim1.0$ mag and $V\sim21.5$ mag, a
red giant branch between $1.0 \lesssim V-I\lesssim1.65$ mag and $17
\lesssim V\lesssim21.5$ mag and a populated red clump at $V\sim19.5$ mag
and $V-I\sim1.0$ mag. The CMD naturally includes also field stars and
stars of the nearby young association NGC 346 (see also Paper I). Part
of the latter is visible for instance as the population of the upper
main sequence.

%\clearpage   
%%%%%%%%%%%%%%%%%%%%%%%%%%%% FIGURE %%%%%%%%%%%%%%%%%%%%%%%%%%%%%%%%%%%%%%%
%%%%%%%%%%%%%%%%%%%%%%%%%%%%%%%%%%%%%%%%%%%%%%%%%%%%%%%%%%%%%%%%%%%%%%%%%%%%

%\clearpage
%%%%%%%%%%%%%%%%%%%%%%%%%%%% FIGURE %%%%%%%%%%%%%%%%%%%%%%%%%%%%%%%%%%%%%%%
\begin{figure*}[t!]
\epsscale{1.}
\plotone{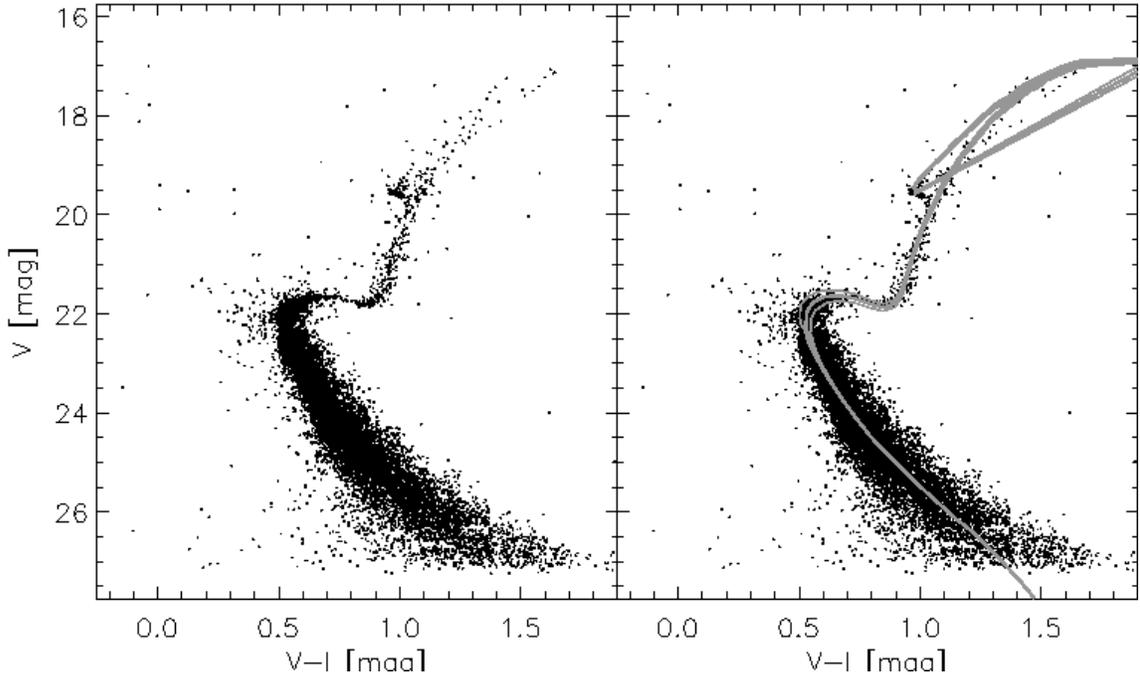}
\caption{{\em Left}: $V$, $V-I$ CMD of BS 90 after the contaminating
stellar population of the general SMC field and the association NGC 346
has been randomly subtracted with the Monte-Carlo method. The
contaminating stars of the upper main sequence, are almost completely
removed. Remaining stars with $21.5 \lesssim V\lesssim22.5$ mag and
$0.2\lesssim V-I\lesssim0.6$ mag may indicate a population of blue  
stragglers. {\em Right}: The same CMD with the applied isochrone models
overplotted.  The best-fitting isochrone, plotted by the middle thick 
line, gives an age of $\tau\simeq4.5$ Gyr. The remaining two models,
plotted with thin lines, correspond to the limiting isochrones that can
also be fitted, and provide the error of 0.5 Gyr in the age estimation.
All three models fit the cluster best for a metallicity
$\rm{[Fe/H]}\simeq-0.72$, an assumed distance modulus of $m-M\simeq18.85$
mag, and a total visual extinction $A_V\sim0.04$ mag. The features of the  
red clump, the red giant branch, the sub-giant branch and the turn-off are
all fitted accurately.}
\label{fig-cmd}
\end{figure*}
%%%%%%%%%%%%%%%%%%%%%%%%%%%%%%%%%%%%%%%%%%%%%%%%%%%%%%%%%%%%%%%%%%%%%%%%%%%%

\subsection{Subtraction of Contaminating Stars}\label{subtraction}

In order to have a `clean' CMD, composed only by stars of the cluster,
a subtraction of all contaminating populations is necessary. Therefore
we applied a Monte Carlo subtraction method to clean the diagram from
stars, that belong to the field and the association.

Naturally, this subtraction might depend on the area chosen to be field. 
Considering that the observed field-of-view (FoV) is no more than 
$5\arcmin \times 5\arcmin$ in size, and that the association NGC~346 
itself also contaminates the stellar population of BS~90, we tested 
several ``empty'' areas within the FoV, which also include stars from the 
association, as potential fields, in order to determine the most realistic 
stellar numbers that should be reduced from the CMD of BS~90. As a 
reference for the stellar contamination we selected an area of 
$75\arcsec\times50\arcsec$ in the south of BS 90 with its center 
$\sim90\arcsec$ away from the center of the cluster.  It contains stars of 
the association, of the general field and also few stars of the cluster 
itself. The number of stars that belong to the cluster is negligible at 
distances of $r>65\arcsec$ from its center, because the stellar surface 
density has almost dropped to the background level of the field at this 
distance (see Fig.~\ref{fig-dpl}). We made the assumption that the 
selected area contains only field stars and stars of the association and 
not stars of BS 90. This leads to a small overestimation of the number of 
stars to be subtracted and causes a slightly higher number of subtractions 
in the area of the sub-giant branch and the main sequence. 

We divided the CMD into a grid and counted the numbers of stars in each 
grid element of the CMD of the reference field. Considering that the 
field-subtraction might also depend on the size of the elements, we tested 
the Monte Carlo subtraction for different grid element sizes, and we found 
that a size of 0.125 mag is the most suitable. Smaller element sizes 
produce poor number statistics, while larger sizes result to rough 
field-subtracted CMDs. Furthermore, we considered the different 
completeness of the two regions by correcting the corresponding stellar 
numbers for every field. Finally, we normalized the stellar numbers to the 
selected area of the cluster. The resulting number of stars which 
contaminates the CMD of BS 90 has been removed. The stars which have been 
excluded have been randomly selected from the corresponding bin of the 
original CMD of Fig.~\ref{fig-cmd_all}. The derived `clean' CMD of BS 90, 
with no contamination by other stellar populations is shown in 
Fig.~\ref{fig-cmd}.

\subsection{The CMD of the Cluster}\label{cmdcluster}

The CMD of Fig.~\ref{fig-cmd} ({\em left}), which corresponds to the CMD 
of the star cluster BS 90 alone shows features of a {\em bona-fide} 
intermediate-age cluster. The diagram reveals a well populated faint main 
sequence, as well as a clear turn-off. Well-defined features are also the 
sub-giant branch (SGB), the red giant branch (RGB) and the red clump (RC). 
The asymptotic giant branch is sparsely populated. These clear 
characteristics of the cluster allow isochrones to be easily used in order 
to define several properties of BS 90. Specifically the SGB, the RGB and 
the RC show no significant differential reddening, which makes the 
application straightforward. We used isochrone models calculated by the 
Padova Group (Girardi et al. 2002) in the VEGA filter system for ACS 
filters.

The best fitting isochrones are shown in the right panel of
Fig.~\ref{fig-cmd} superimposed on the CMD. The chosen isochrone fits
very well the turn-off, RC, SGB and RGB. These well-defined features of
BS 90 allow us to make an accurate estimation of the characteristic properties
of the cluster:

\begin{itemize}
\item[{\em (i)}] {\em Age and Metallicity}\\
The best-fitting isochrones from the used model grid give an age of
$4.5\pm0.5$ Gyr and a metallicity of $Z=0.004\pm0.001$. This metallicity
leads through the relation ${\rm [Fe/H]}= 1.024\cdot \log{Z}+1.739$
(Bertelli et al. 1994) to an iron content for BS 90 of
$\rm{[Fe/H]}=-0.72^{+0.10}_{-0.13}$.

\item[{\em (ii)}] {\em Reddening and Distance}\\
From our isochrone fit we found a color excess of only $E(V-I)\simeq
0.016$ mag. Using the relation between the extinction in $V$, $A_V$, and
the color excess $E(V-I)$, derived by Mackey \& Gilmore (2003c), we
obtained a total visual extinction of $A_V\simeq 0.038$ mag. We assumed
a normal extinction law with a total-to-selective absorption of
$R_V=3.1$ (Binney \& Merrifield 1998), which gives $E(B-V)\simeq 0.021$
mag.

The distance modulus of BS 90 has been determined to
$m-M\simeq18.85\pm0.1$ mag, which corresponds to a distance of
$58.9\pm0.45$ kpc.

\end{itemize}

%\clearpage
%%%%%%%%%%%%%%%%%%%%%%%%%%%% FIGURE %%%%%%%%%%%%%%%%%%%%%%%%%%%%%%%%%%%%%%%
\begin{figure*}[t!]
\epsscale{1.}   
\plotone{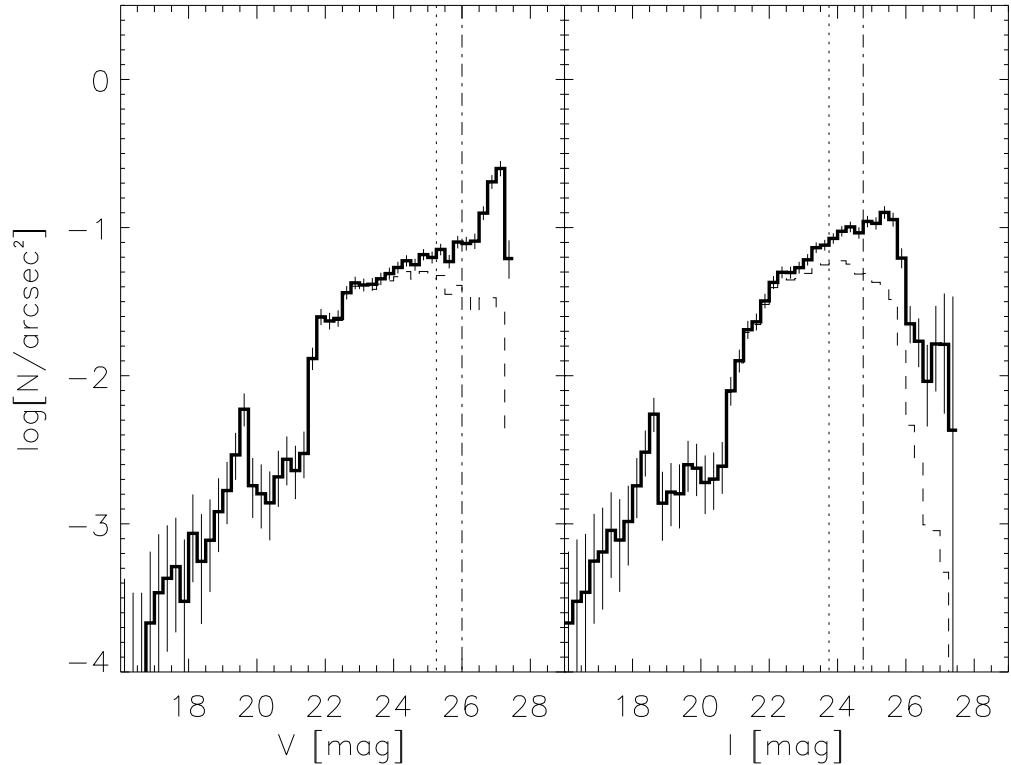}
\caption{The field-subtracted Luminosity Functions in $V$ ({\em left}) 
and $I$ ({\em right}) of the cluster within a radius $2\farcm8$. The
features of the red clump around $V\sim 19$ mag and $I\sim 18.5$ mag,
and the turn-off at $V\sim 21.5$ mag and $I\sim 21$ mag are clearly
seen. The increase of the stellar density for $V\gtrsim 26.5$ mag is
probably due to the very low completeness in our data. The vertical
dotted line represents the 70\% completeness limit, while the vertical
dash-dotted line represents the 50\% completeness limit. The errors 
represent Poisson statistics. The dashed line shows the luminosity
function before it has been corrected for incompleteness.}
\label{fig-lf}
\end{figure*}
%%%%%%%%%%%%%%%%%%%%%%%%%%%%%%%%%%%%%%%%%%%%%%%%%%%%%%%%%%%%%%%%%%%%%%%%%%%

\subsubsection{Blue Straggler Candidates}

As shown in the CMD of Fig.~\ref{fig-cmd}, after the field subtraction
there are still few stars which remain located a bit above the turn-off
and to the blue. These stars appear to be main sequence stars with
masses larger than expected for stars of the cluster at its turn-off
age. The region of the CMD covered by these stars is between
$V\simeq22.5$ and $21$ mag, comparable to the region on the CMD, where
blue straggler stars are found in the LMC cluster ESO 121-SC03 (Mackey,
Payne \& Gilmore 2006). The spatial distribution of these candidate blue
stragglers in BS 90 shows that they are concentrated in the center of the
cluster. This supports the suggestion that these stars represent indeed a
blue straggler population, since such stars are centrally concentrated. A
suggested explanation for the Blue Stragglers phenomenon is mass
exchange in binary systems and merging processes, which increase the mass
of one star and, therefore, it shines brighter and is bluer than expected
for its age.

\section{Luminosity Function and Mass Function}\label{lfmf}

\subsection{Luminosity Function}\label{lf}

We constructed the $V$- and $I$-band luminosity functions (LF) of the
cluster, which are presented in Fig.~\ref{fig-lf}. For the construction
of the LF we considered the CMD of the cluster shown in
Fig.~\ref{fig-cmd}, without the contamination by the populations of the
association and the background field within a radius $2\farcm8$ from the
center of the cluster (comparable to $r_{\rm t}$). We counted the stars
in magnitude bins of $0.25$ mag and corrected the numbers for
incompleteness.  The LF has been normalized to the considered surface of
$6.5\ arcmin^2$. The completeness corrected LF is shown with thick
solid lines. The vertical dotted lines in Fig.~\ref{fig-lf}, which
correspond to $V=25.75$ mag and $I=24.5$ mag respectively, represent the
50\% completeness limit. The vertical dashed-dotted line at $V=25$ mag
and $I=23.75$ mag respectively corresponds to the 70\% completeness
limit. The dashed line shows the LF before the completeness correction.

In the LFs of Fig.~\ref{fig-lf} different features of the cluster can be
clearly identified. The red clump is visible as the increase in
stellar density around $V=19.5$ mag. The turn-off is also
distinguishable at $V\simeq 21.5$ mag. For fainter stars the LF turns
into a linearly increasing function, showing the main sequence stars of
the cluster. In the case of the $I$-LF we observe similar features.
Around $I=18.5$ mag we identify the red clump and we see the turn-off at
$I\simeq 20.75$ mag. For stars with $20.75 \lesssim I\lesssim22$ mag the
linear increase of the LF is steeper than for $22 \lesssim
I\lesssim24.5$ mag. The latter range is composed of stars of the main
sequence.

\subsection{Mass Function}\label{mf}

The IMF is a function which describes the mass distribution of a newly
formed stellar population. The stellar mass function 
is changed under the influence of stellar, as well as dynamical evolution 
that affects entire stellar systems such as star clusters (Baumgardt \& Makino 2003). 
The present-day mass function (PDMF) is the result of such evolutionary effects. 
In general, the IMF can be described as a power law (Scalo 1986):
\begin{equation}
\xi(\log{\rm m})\propto m^{\Gamma}
\label{imf}
\end{equation}
and represents the distribution of stellar masses for a given stellar system
at the time of their formation. The index $\Gamma$ is its power-law slope
\begin{equation}
\Gamma=\frac{d\log{\xi(\log{\rm m})}}{d\log{\rm m}}.
\label{gamma}
\end{equation}

Salpeter (1955) found a value for the slope of the IMF in the solar
neighborhood, often taken as a reference, of $\Gamma=-1.35$. In the same
manner as for the IMF, we describe the slope of the PDMF as
\begin{equation}
a=\frac{d\log{N(\log{\rm m})}}{d\log{\rm m}}\ ,
\label{alpha}
\end{equation}
where $N(\log{\rm m})$ is the PDMF. This slope is given by the linear
relation between the logarithmic mass intervals and the corresponding
number of stars (in logarithmic scale). In order to obtain the PDMF of a
cluster, knowledge of the mass-luminosity relation (MLR) is required. In
the case of BS 90, and since a single age model fits almost perfectly
the cluster, we used the MLR taken from the 4.5 Gyr isochrone from
Girardi et al. (2002) and ascribed the present mass to each star which
belongs to our stellar sample of the cluster.  To construct the PDMF of
BS 90, we counted stars in logarithmic mass intervals and normalized
their numbers to the considered area. We corrected these numbers for
incompleteness. 

\begin{figure}[t!]
\epsscale{1.2}
\plotone{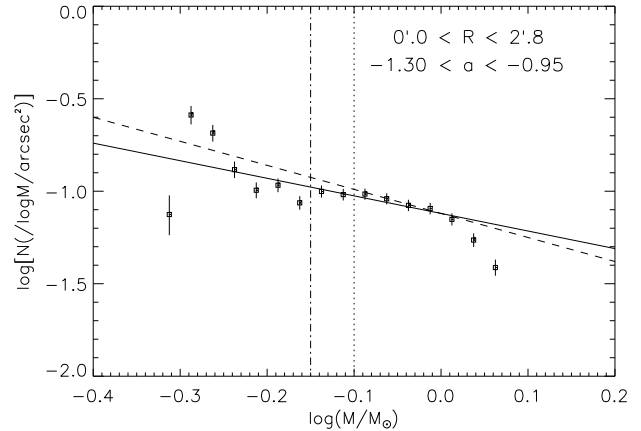}
\caption{The field-subtracted, completeness corrected present-day mass
function of the cluster within a radial distance of $2\farcm8$ from its
center. The steep increase to the lower mass end ($\log{(M/{\rm
M}_{\odot})} <-0.25$) is probably due to the very low completeness. The
vertical dotted line represents the 70\% completeness limit while the
vertical dash-dotted line represents the 50\% completeness limit. The  
best linear fit for data with completeness higher than 50\%,
corresponding to a slope of $a\simeq-0.95$ is plotted with a solid   
line. The dashed line shows the linear fit to the mass function for
stars with completeness higher than 70 \% with a slope of $a\simeq-1.30$.}
\label{fig-mf}
\end{figure}

The constructed PDMF is shown in Fig.~\ref{fig-mf}, where an almost
linear increase to lower masses can be seen.  The low- and high-mass
ends of the PDMF show a steep increase and decrease respectively. In the
case of the low-mass end this is probably due to the completeness
corrections, while for the high-mass end stellar evolution is possibly
responsible. The 50\% completeness limit corresponds to a mass of
$M\sim0.7\ \rm{M_{\odot}}$ and is represented by the vertical
dash-dotted line. The 70\% completeness limit is shown with a vertical
dotted line and represents a lower mass limit of $M\sim0.8\
\rm{M_{\odot}}$. The straight solid line shows the fit to the mass
function above the 50\% completeness limit, without the two highest mass
points. The fit gives a single power-law slope of $a=-0.95\pm0.14$, a
bit flatter than Salpeter's IMF, whereas the fit to the mass function 
above the 70\% completeness limit reveals a Salpeter-like slope of 
$a=-1.30\pm0.14$. It is represented by the dashed line.

Kroupa (2002) notes that theoretical examinations of the dynamics of massive
and long-lived globular clusters ($N\gtrsim 10^5$ stars) reveal that
the global PDMF is similar to the PDMF at distances near the half-mass
radius.  The PDMF inward and outward the half-mass radius becomes
flatter and steeper, respectively, due to dynamical mass segregation.
Dynamical evolution of a cluster also flattens the global PDMF, since
evaporation is stronger for stars with lower masses than for high-mass
stars (Vesperini \& Heggie 1997).

%\clearpage
%%%%%%%%%%%%%%%%%%%%%%%%%%%% FIGURE %%%%%%%%%%%%%%%%%%%%%%%%%%%%%%%%%%%%%%%
%%%%%%%%%%%%%%%%%%%%%%%%%%%%%%%%%%%%%%%%%%%%%%%%%%%%%%%%%%%%%%%%%%%%%%%%%%%

\section{Mass segregation in BS 90}\label{segregation}

\begin{figure*}[t!]
\epsscale{1.}
\plotone{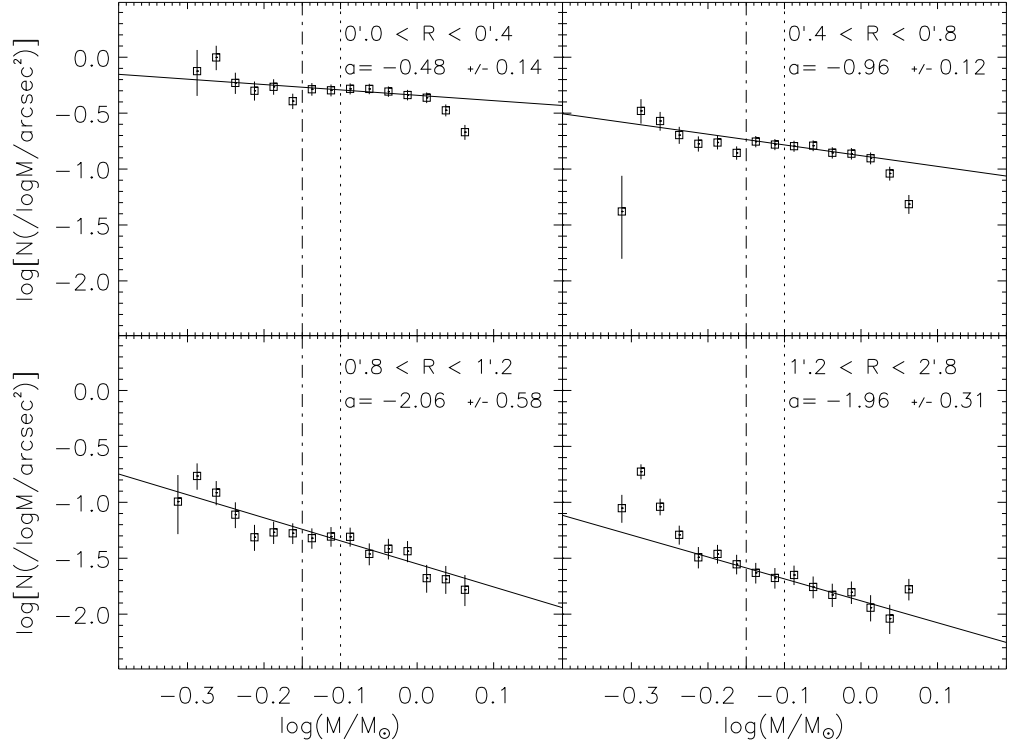}
\caption{Mass functions of BS 90 for four selected radial distances
($0\farcm0-0\farcm4$, $0\farcm4-0\farcm8$, $0\farcm8-1\farcm2$,
$1\farcm2-2\farcm8$). The dotted and dashed-dotted lines represent the
70\% and 50\% completeness limit respectively. The linear fits to the
MFs are represented by the solid line in each panel. The mass function
becomes steeper for larger radial distances from the center of the
cluster indicating that mass segregation takes place.}
\label{fig-mseg}
\end{figure*}

The phenomenon of mass segregation describes the concentration of
high-mass stars towards the center of a star cluster. This phenomenon
could be either of primordial origin or the result of the dynamical
evolution of the cluster. The latter effect leads the cluster to a state of
energy equipartition between its stars. Therefore, stars with lower
masses will have higher velocities and thus larger orbits. On the other
hand, stars with higher masses have smaller orbits and
concentrate towards the center. Indeed, stellar mass functions (MFs) at
the outer parts of star clusters indicate the existence of more low-mass
stars at larger radii than in the central region of the cluster (de
Grijs et al. 2002). A powerful diagnostic for the detection of the
phenomenon of mass segregation in a star cluster is its MF, which is expected to be radially
variable if segregation is present (e.g. Gouliermis et al. 2004).
Therefore, in order to see if BS 90 is mass segregated or not, we derived
its MF for annuli at different radial distances from the center of the
cluster. We derived the different MFs as described in \S~\ref{mf} for
four selected radial distances $0\farcm0-0\farcm4$, $0\farcm4-0\farcm8$,
$0\farcm8-1\farcm2$ and $1\farcm2-1\farcm6$. The resulting MFs are shown
in Fig.~\ref{fig-mseg}. The MF of the innermost region has a slope of
$a=-0.48\pm0.14$. The MF of the second annulus, which includes the
half-light radius, is comparable to the MF inside $2\farcm8$ shown in
Fig.~\ref{fig-mf} and has a slope of $a=-0.96\pm0.12$. The slopes of
the outer regions are found to be steeper with $a=-2.06\pm0.58$ for
$0\farcm8-1\farcm2$ and $a=-1.96\pm0.31$ for $1\farcm2-2\farcm8$.  The
difference of the MF slope being steeper outwards, gives clear evidence
of mass segregation in BS 90. In the following section, we will explore the
driving mechanism of this phenomenon.

%\clearpage
%%%%%%%%%%%%%%%%%%%%%%%%%%%% FIGURE %%%%%%%%%%%%%%%%%%%%%%%%%%%%%%%%%%%%%%%
%%%%%%%%%%%%%%%%%%%%%%%%%%%%%%%%%%%%%%%%%%%%%%%%%%%%%%%%%%%%%%%%%%%%%%%%%%%

\subsection{Half-Mass Radius and Dynamical Time-Scale}\label{halfmass}

As mentioned above, the phenomenon of mass segregation may be
primordial, due to the formation of the most massive stars in the
central regions of the cluster, or an evolutionary effect due to the
dynamical relaxation of the cluster.  The latter is achieved via
two-body encounters, which lead to a quasi-Maxwellian equilibrium in
the interior of the cluster (Lightman \& Shapiro 1978).  The tendency of
the inner regions of the cluster to reach thermal equilibrium leads to
energy equipartition among different stellar mass groups and
consequently to mass stratification. The time-scale in which such
equipartition is accomplished is given roughly by the relaxation time,
$t_{\rm rl}$, of the cluster (e.g. Spitzer 1975). Hence, the relaxation
time can indicate whether a case of mass segregation is of dynamical
origin (when the evolution time of the stellar system is longer than the
relaxation time) or not. This time-scale can be expressed as (Binney \&
Tremaine 1987):
\begin{equation}
t_{\rm rl}=\frac{6.63\times10^8}{\ln{0.4N}}\left(\frac{M}{10^5{\rm M}_{\odot}}\right)^{1/2}\left(\frac{{\rm M}_{\odot}}{m_{\star}}\right)\left(\frac{\mathit{R}}{\rm pc}\right)^{3/2}~~~{\rm yr}
\label{eq-rel}
\end{equation}
where $M$ is the total mass in a certain characteristic radius $R$,
$m_{\star}$ is a characteristic mass (median mass of the observed
stellar distribution) and $N$ is the corresponding number of stars of
the system. The half-mass radius corresponds more or less to the
half-light radius of the cluster. The half-light radius of BS 90 derived
from its surface brightness profile (\S~\ref{halflight}) is found to be equal
to $r_{\rm h}=9.42\pm0.5$ pc. Following Eq.~\ref{eq-rel}, we derive the
relaxation times, corresponding to the half-light and core radii, after
we estimated the corresponding total and characteristic mass and the
total number of stars within these radii. We found a half-mass
relaxation time of $t_{\rm rl,h}=0.95\pm0.06$ Gyr and a core radius
relaxation time of $t_{\rm rl,c}=0.33^{+0.06}_{-0.04}$ Gyr. From these
timescales, which are much smaller than the evolutionary age of the
cluster, we easily conclude that BS 90 is a dynamically relaxed
spherical cluster, which thus exhibits the phenomenon of dynamical mass
segregation.

\section{Chemical Evolution History of the SMC}\label{ceh-smc}

The chemical evolution history of a galaxy is understood in terms of the
age-metallicity relation of both field and cluster stars. The chemical
evolution histories in the MCs exhibit distinct features in comparison
to this of the disc of the Milky Way (Westerlund 1997). Several
models have been developed for their explanation, assuming bursts with
IMFs steeper than Salpeter's IMF to produce subsolar metallicities (e.g.
Russell \& Dopita 1992; Tsujimoto et al. 1995), or postulating selective
stellar outflows associated with star formation bursts (e.g. Pilyugin
1996). In contrast, the model of chemical evolution for the MCs
developed by Pagel \& Tautvai\v{s}ien\'{e} (1998) assumes that the
Clouds have been built up by gradual infall of unprocessed material,
linear laws of star formation, yields and time delays identical to those
for the solar neighborhood, and a non-selective wind proportional to
the star formation rate. These models represent chemical enrichment
assuming a burst-like star formation (`bursting' model) and a continuous
star formation rate over the entire lifetime of the SMC (`smooth', or simple
`closed box' model). 

%\clearpage
%%%%%%%%%%%%%%%%%%%%%%%%%%%%%%%%%%%%%%%%%%%%%%%%%%%%%%%%%%%%%%%%%%%%%%%%%%
\begin{table*}
\caption{Collective Data on ages and metallicities of star clusters in the SMC. \label{table1}}
\begin{center}
\scriptsize{
\begin{tabular}{lccrrrrrrc}
\hline
Name & 		RA & 	DEC & 	Age   &     	 &         &    Metal.   &	      &  & 	     References\\
&\multicolumn{2}{c}{J2000}& 	(Gyr)   &    + 	 &    $-$     &   [Fe/H]   &	 +     & $-$ & 	    \\
\hline
\hline
NGC 121  & 	00:26:49 & 	$-$71:32:10 & 	12.00 &    2.00  &    2.00  &    $-$1.19  &	0.12  &   0.12   &    Da Costa \& Hatzidimitriou (1998)\\
 &  &	 &			 					11.90  &    1.30   &    1.30   &    $-$1.71  &	0.10  &   0.10   &   Mighell et al. (1998)\\
 &  &	 &								12.00  &    5.00   &    5.00   &    $-$1.20  &	0.32  &   0.32   &    de Freitas Pacheco et al. (1998)\\
NGC 152  &	00:32:56 & $-$73:06:57   & 	1.40   &    0.20   &    0.20   &    $-$0.94  &	0.15  &   0.15   &   Crowl et al. (2001)\\
 & 	 & 	 & 		 				 	 	1.90   &    0.50   &    0.50   &    $-$0.80  &	0.30   &   0.30    &   Da Costa \& Hatzidimitriou (1998)\\
NGC 176  & 	00:35:59 & 	$-$73:09:57  & 	0.20   &    0.20   &    0.10   &    $-$0.60   &	-  &   -   &   Mackey \& Gilmore (2003b)\\
NGC 339  & 	00:57:48 & 	$-$74:28:00  & 	4.00   &    1.50  &    1.50  &    $-$1.19  &	0.10  &   0.10   &   Da Costa \& Hatzidimitriou (1998)\\
 &  & 	 & 				   	6.30   &    1.30   &    1.30   &    $-$1.50  &	0.14  &   0.14   &   Mighell et al. (1998)\\
 &  & 	 & 				   	2.00   &    0.70   &    0.70   &    $-$0.70  &	0.22  &   0.22   &   de Freitas Pacheco et al. (1998)\\
NGC 361  & 	01:02:13 & 	$-$71:36:16  & 	8.10   &    1.20   &    1.20   &    $-$1.45  &	0.11  &   0.11   &   Mighell et al. (1998)\\
NGC 411  & 	01:07:56 & 	$-$71:46:05  & 	1.40   &    0.20   &    0.20   &    $-$0.68  &	0.07  &   0.07   &   Alves \& Sarajedini (1999)\\
 &  & 	 & 		   			1.30   &    0.50   &    0.30   &    $-$0.70  &	0.22  &   0.22   &   de Freitas Pacheco et al. (1998)\\
 &  & 	 & 		   			1.50   &    0.30   &    0.30   &    $-$0.70   &	0.20   &   0.20    &   Piatti et al. (2005b)\\
 &  & 	 & 		   			1.80   &    0.30   &    0.30   &    $-$0.84  &	0.30   &   0.30    &   Da Costa \& Hatzidimitriou (1998)\\
NGC 416  & 	01:07:59 & 	$-$72:21:20  & 	6.90   &    1.10   &    1.10   &    $-$1.44  &	0.12  &   0.12   &   Mighell et al. (1998)\\
 &  & 	 & 		   			4.00   &    1.50   &    1.50   &    $-$0.80  &	0.23  &   0.23   &   de Freitas Pacheco et al. (1998)\\
NGC 419  & 	01:08:19 & 	$-$72:53:03  & 	4.00   &    1.50   &    1.50   &    $-$0.60  &	0.21  &   0.21   &   de Freitas Pacheco et al. (1998)\\
 & & 	 & 		 	  		1.20   &    0.50   &    0.50   &    $-$0.70   &	0.30   &   0.30    &   Da Costa \& Hatzidimitriou (1998)\\
NGC 458  & 	01:14:53 & 	$-$71:32:59  & 	0.20  &    0.10    &    0.10  &    $-$0.23  &	0.10   &   0.40    &   Da Costa \& Hatzidimitriou (1998)\\
 &  & 	 & 		   			0.13  &     0.06  &    0.06    &    $-$0.23  &	0.10   &   0.40    &   Piatti et al. (2005b)\\
L1   	 &	00:04:00 & 	$-$73:28:00  & 	10.00 &    2.00  &    2.00  &    $-$1.01  &	0.11  &   0.11   &   Da Costa \& Hatzidimitriou (1998)\\
 &  & 	 & 		   			9.00   &    1.00   &    1.00   &    $-$1.35  &	0.08  &   0.08   &   Mighell et al. (1998)\\
L4  	 & 	00:21:27 & 	$-$73:44:55  & 	3.10   &    0.90   &    0.90   &    $-$0.90   &	0.20   &   0.20    &   Piatti et al. (2005a)\\
L5 	  & 	00:22:40 & 	$-$75:04:29  & 	4.10   &    0.90   &    0.90   &    $-$1.20  &	0.20  &   0.20   &   Piatti et al. (2005a)\\
 &  & 	 & 		   			3.00   &    1.50   &    1.50   &    $-$1.10   &	0.20   &   0.20    &   Piatti et al. (2005b)\\
L6/K4   & 	00:23:04 & 	$-$73:40:11  & 	3.30   &    0.90   &    0.90   &    $-$0.90   &	0.20   &   0.20    &   Piatti et al. (2005a)\\
L7/K5   & 	00:24:43 & 	$-$73:45:18  & 	2.00   &    0.20   &    0.20   &    $-$0.60   &	0.20   &   0.20    &   Piatti et al. (2005a)\\
 &  & 	 & 		   			1.20 &    0.50   &    0.50   &  	 $-$0.50   &	0.20   &   0.20    &   Piatti et al. (2005b)\\
L11/K7 	 & 	00:27:45 & 	$-$72:46:53  & 	3.50  &    1.00  &    1.00  &    $-$0.81  &	0.13  &   0.13   &   Da Costa \& Hatzidimitriou (1998)\\
 &  & 	 & 		   			3.50   &    0.50   &    0.50   &    $-$1.00   &	-  &	-	 &   Piatti et al. (2005b)\\
L19 	 & 	00:37:42 & 	$-$73:54:30  & 	2.10   &    0.20   &    0.20   &    $-$0.75  &	0.20   &   0.20    &   Piatti et al. (2005a)\\
L27 	 & 	00:41:24 & 	$-$72:53:27  & 	2.10   &    0.20   &    0.20   &    $-$1.30 	&	0.30   &   0.30    &   Piatti et al. (2005a)\\
L32 	 &	00:47:24 & 	$-$68:55:10  & 	4.80  &    0.50  &    0.50  &    $-$1.20   &	0.20   &   0.20    &   Piatti et al. (2001)\\
L38 	 &	00:48:50 & 	$-$69:52:11  & 	6.00  &    0.60  &    0.60  &    $-$1.65  &	0.20   &   0.20    &   Piatti et al. (2001)\\
L113 	 &	01:49:30 & 	$-$73:43:40  & 	6.00  &    1.00  &    1.00  &    $-$1.17  &	0.12  &   0.12   &   Da Costa \& Hatzidimitriou (1998)\\
 &  & 	 & 		   			5.30   &    1.30   &    1.30   &    $-$1.24  &	0.11  &   0.11   &   Mighell et al. (1998)\\
L116     & 	01:55:33 & 	$-$77:39:16  & 	2.80  &    1.00  &    1.00  &    $-$1.10   &	0.20   &   0.20    &   Piatti et al. (2001)\\
K3  & 	 & &		   					6.00   &    1.30   &    1.30   &    $-$1.16  &	0.09  &   0.09   &   Mighell et al. (1998)\\
 &  & 	 & 		   						9.00  &    2.00  &    2.00  &    $-$0.98  &	0.12  &   0.12   &   Da Costa \& Hatzidimitriou (1998)\\
 &  & 	 & 		   						3.50   &    1.50   &    1.50   &    $-$1.00  &	0.28  &   0.28   &   de Freitas Pacheco et al. (1998)\\
 &  & 	 & 		 		 			 	7.00   &    1.00   &    1.00   &    $-$1.20  &	0.20   &   0.20    &   Piatti et al. (2005b)\\
K28 	 & 	00:51:42 & 	$-$71:59:52  & 	2.10   &    0.50   &    0.50   &    $-$1.20  &	0.20   &   0.20    &   Piatti et al. (2001)\\
 &  & 	 & 		 		  	1.50   &    0.60   &    0.60   &    $-$1.00   &	0.20   &   0.20    &   Piatti et al. (2005b)\\
K44 	 & 	01:02:04 & 	$-$73:55:31  & 	3.10  &    0.80  &      0.80  &    $-$1.10   &	0.20   &   0.20    &   Piatti et al. (2001)\\
HW 47    & 	01:04:04 & 	$-$74:37:09  & 	2.80   &    0.90   &    0.90   &    $-$1.00   &	0.40   &   0.40    &   Piatti et al. (2005a)\\
HW 84    & 	01:41:28 & 	$-$71:09:58  & 	2.40   &    0.20   &    0.20   &    $-$1.20	&	0.40   &   0.40    &   Piatti et al. (2005a)\\
HW 86    & 	01:42:22 & 	$-$74:10:24  & 	1.60   &    0.20   &    0.20   &    $-$0.75  &	0.40   &   0.40    &   Piatti et al. (2005a)\\
BS 121   & 	01:04:22 & 	$-$72:50:52  & 	2.30   &    0.20   &    0.20   &    $-$1.20   &	0.40   &   0.40    &   Piatti et al. (2005a)\\
BS90     & 	00:59:05 & 	$-$72:09:10  & 	4.30   &    0.10  &     0.10  &     $-$0.84  &	-   &   -	 &   Sabbi et al. (2007)\\
         &               &                   & 	4.47   &    0.55  &     0.49  &     $-$0.71  &	0.10   &   0.13	 &   This Paper\\
\hline
\hline
\end{tabular}
}
\end{center}
\end{table*}
%%%%%%%%%%%%%%%%%%%%%%%%%%%%%%%%%%%%%%%%%%%%%%%%%%%%%%%%%%%%%%%%%%%%%%%%%%

\begin{figure}[t!]
\epsscale{1.25}
\plotone{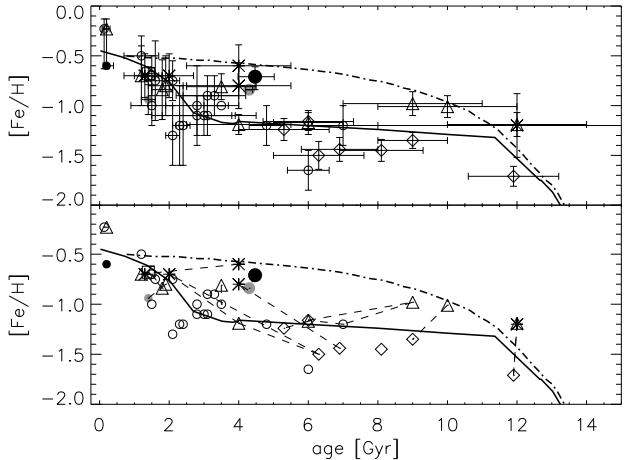}
\caption{Age-metallicity relation of the cluster population in the SMC.
The data are obtained from Crowl et al. (2001; small grey dot), Piatti
et al. (2001, 2005a,b; small open circles), Mighell et al. (1998;
diamonds), Da Costa \& Hatzidimitriou (1998; triangles), de Freitas
Pacheco (1998; asterisks), Mackey \& Gilmore (2003b; small black dot),
Alves \& Sarajedini (1999; square), Sabbi et al. (2007; large grey dot).
Our measurements for BS 90 are presented as the large black dot.  In
both panels the bursting (solid curve) and the smooth (dotted-dashed
curve) chemical evolution models by Pagel \& Tautvai\v{s}ien\'{e} (1998)
are also shown. {\em Upper panel}: The values for all studied SMC
clusters (shown in Table~\ref{table1}) are plotted with their
corresponding errors.  {\em Lower panel}: The same values from
Table~\ref{table1} are shown without their errors. The dashed lines   
connect the values from estimates derived from different studies for the
same clusters, exhibiting, thus, the discrepancy in the results of
different methods.}
\label{fig-amr}
\end{figure}

In order to reproduce the age-metallicity relation for individual clusters 
in the SMC and add our results on BS 90, we collected a large set of data 
available on SMC clusters, concerning their metallicities and ages. These 
data are given in Table~\ref{table1}. We plot the age-metallicity relation 
for the SMC clusters in Fig. ~\ref{fig-amr}. The models published by Pagel 
\& Tautvai\v{s}ien\'{e} (1998) are also displayed with a solid (`bursting' 
model) and a dash-dotted line (`smooth' model). The upper panel of 
Fig.~\ref{fig-amr} shows ages and metallicities with the corresponding 
errors for all star clusters collected in Table~\ref{table1}. The lower 
panel is displayed without the errors, but with a set of dashed lines, 
which show the connections of points derived from different studies for 
the same clusters. The latter shows that both the `bursting' and the 
`smooth' model can be supported by selecting certain clusters and hence an 
unambiguous conclusion is not possible yet. Moreover, it should be noted 
that the data given in Table~\ref{table1} consist of an inhomogeneous set 
of abundances, perhaps washing away any real effects. The transformation of 
all these results into a common spectroscopic scale would certainly 
provide a more clear picture. In any case if we assume that the `bursting' 
model of Pagel \& Tautvai\v{s}ien\'{e} (1998) represents the chemical 
evolution history of SMC, then our measurements reveal a relatively high 
metallicity for the epoch during which BS 90 was formed. These values seem 
to fit better the `smooth' model, which represents a continuous mode of 
chemical enrichment.

%\clearpage
%%%%%%%%%%%%%%%%%%%%%%%%%%%% FIGURE %%%%%%%%%%%%%%%%%%%%%%%%%%%%%%%%%%%%%%%
%%%%%%%%%%%%%%%%%%%%%%%%%%%%%%%%%%%%%%%%%%%%%%%%%%%%%%%%%%%%%%%%%%%%%%%%%%%
%Smooth model represents a continuous mode of enrichment is shown by the
%dotted line.
%%%%%%%%%%%%%%%%%%%%%%%%%%%%%%%%%%%%%%%%%%%%%%%%%%%%%%%%%%%%%%%%%%%%%%%%%%%

Indeed, our estimated age and metallicity for BS 90 places the cluster in
Fig.~\ref{fig-amr} closer to the `smooth' model (large black dot), in
agreement with the estimations of Sabbi et al. (2007) (large grey dot).
According to the `bursting' model an iron abundance of
[Fe/H]$\approx-1.2$ corresponds to the time around 4.5 Gyr ago. Our
derived value for BS 90 is about $0.5$ dex higher. On the other hand the
`smooth' model requires an iron content of [Fe/H]$\approx-0.6$ at $\sim
4.5$ Gyr, much closer to our derived value of [Fe/H]$\approx-0.7$. A
simple `closed box' model is also supported by the results of Kayser et
al. (2006) and consistent with the enrichment history of the outer SMC
field which has been derived by Dolphin et al. (2001). With its age and
metallicity BS 90 adds an important point to the age-metallicity
relation, which favors a continuous mode of chemical enhancement in the
SMC, but the chemical evolution history of this galaxy still remains
elusive and seems to be more complicated than reflected by the
proposed models. Fig.~\ref{fig-amr} also reveals an increasing number of
clusters for the last $\sim 3$ Gyr.  This coincides with the increase of
cluster formation in the LMC and is consistent with a close encounter
between the Magellanic Clouds and Milky Way as proposed by Bekki et al.
(2004).  The strong tidal interactions could certainly trigger cluster
formation.

\section{Summary}\label{sum}

In this paper we present our detailed photometric analysis of the
intermediate age spherical star cluster BS 90 in the SMC.

We analyzed the stellar surface density profile of BS 90, following the
models described by King (1962) and Elson, Fall \& Freeman (1987).  We
derived almost the same value for the core radius of $r_{\rm
c}\simeq5.5$ pc with both models. This value places BS 90 in the $r_{\rm
c}$-age relation among the SMC clusters that do not seem to evolve
isolated. From the model of Elson, Fall \& Freeman (1987), we found a
stellar surface density which, at larger radii, decreases outwards with
a power-law slope of $\gamma\simeq3.68$. The application of King's model
delivered a tidal radius of $r_{\rm t}\simeq53.7$ pc and a concentration
parameter $c\simeq1.0$, which seems to be relatively low, compared to other
SMC clusters (McLaughlin \& van der Marel 2005).

We counted stars in different magnitude intervals in order to derive the 
LF of the cluster. The derived luminosity function
of BS 90 reflects features of the cluster like the red clump as a higher
stellar density around $V=19$ mag and $I=18.5$ mag, respectively, and the
turn-off at $V=22$ mag and $I=20.75$ mag, respectively. The PDMF is a
linearly decreasing function up to the high mass end of the cluster. We
derived a slope of the PDMF of $a\simeq -0.95$ for a completeness in our
photometry higher than 50 \% and of $a\simeq -1.30$ above the 70\% 
completeness limit. Consequently, the PDMF of BS 90 has a slope between 
the one derived by Salpeter (1955) for the IMF in the solar neighborhood, 
and a shallower slope, comparable to earlier derived slopes from studies 
on Magellanic Cloud clusters (e.g. Kerber \& Santiago 2006; Gouliermis 
et al. 2004). The difference from the Salpeter slope can also be explained 
by the fact, that unresolved binaries flatten the PDMF (Kerber \& Santiago 
2006) or that stars escaped from the cluster (Meylan \& Heggie 1997).

We investigated the radial dependence of the mass function slope to
check whether BS 90 is mass segregated or not. We found that indeed the
mass function becomes steeper at larger distances from the center of the
cluster. The radial dependence of the slope indicates a central
concentration of the massive stars and hence mass segregation. We also
confirm previous claims that the PDMF around the half-mass radius is
comparable to the overall PDMF since it is the region which is least
affected by mass segregation (Kroupa 2002; Vesperini \& Heggie 1997). 
Within the half-light radius of $\simeq 9.4$ pc, we obtained a half-mass
relaxation time of about $0.95$ Gyr and therefore we conclude that the
mass segregation is due to the dynamical evolution of BS 90, since the
age of the cluster is older than the derived half-mass relaxation time.
Mass segregation provides another possible reason for a flattening of
the mass function which may occur during the evolution of the cluster.

Isochrone models fitting to the constructed CMD (Fig.~\ref{fig-cmd}) of
BS 90 revealed an age of about $4.5$ Gyr, which characterizes this
cluster as one of the very rare star clusters, formed in a possible
quiescent period of cluster formation in the SMC (Rafelski \& Zaritsky 2005).
There are only few clusters in the range between 4 and 10
Gyr. The applied isochrone yields a metallicity of $\rm{[Fe/H]}=-0.72$. 
The total visual extinction has been found to be $A_{V}\sim 0.04$ mag. The
low value of the total visual extinction combined with the large amount
of gas in the general region and the absence of indications of
differential reddening in the CMD of BS 90 suggests that the cluster is
probably located in front of the association NGC 346. The distance of BS
90 has been estimated to $\simeq$ 58.9 kpc.

The location of BS 90 in the age-metallicity plot shows that this cluster 
fits better to a simple `smooth' model, which represent a continuous 
chemical enhancement for the SMC, rather than a `bursting' model for the 
chemical evolution history of the galaxy, in agreement with recent results 
on other SMC clusters (Kayser et al. 2006). However, inconsistent 
metallicity estimation methods for different SMC clusters provide an 
inhomogeneous set of results, that should be transfered onto a common 
abundance scale, so that an unambiguous picture of the chemical evolution 
history of SMC can be derived. Furthermore, given that the SMC is part of an interacting system between the MCs and the Milky Way, the chemical evolution history of this galaxy is probably far more complicated than what suggested by a simple `closed-box' model.

\acknowledgments

Dimitrios A. Gouliermis\ acknowledges the support of the German Research
Foundation (Deu\-tsche For\-schungs\-ge\-mein\-schaft - DFG) through the
individual grant 1659/1-1. Boyke Rochau would like to thank Sascha P.
Quanz and Alessandro Berton for their helpful suggestions. This paper is
based on observations made with the NASA/ESA Hubble Space Telescope,
obtained from the data archive at the Space Telescope Science Institute.
STScI is operated by the Association of Universities for Research in
Astronomy, Inc. under NASA contract NAS 5-26555. 

%\reference{} 
%%%%%%%%%%%%%%%%%%%%%%%%%%%  BIBLIOGRAPHY  %%%%%%%%%%%%%%%%%%%%%%%%%%%%%%%
%%%%%%%%%%%%%%%%%%%%%%%%%%%%%%%%%%%%%%%%%%%%%%%%%%%%%%%%%%%%%%%%%%%%%%%%%%
%\newpage

%%%%%%%%%%%%%%%%%%%%%%%%%%%%%%%%%%%%%%%%%%%%%%%%%%%%%%%%%%%%%%%%%%%%%%%%%%


\begin{references}

\reference{} Alves, D.~R., \& Sarajedini, A.\ 1999, \apj, 511, 225
\reference{} Baumgardt, H., \& Makino, J.\ 2003, \mnras, 340, 227
\reference{} Bekki, K., Couch, W.~J., Beasley, M.~A., Forbes, D.~A., Chiba, M., \& Da Costa, G.~S.\ 2004, \apjl, 610, L93
\reference{} Bertelli, G., Bressan, A., Chiosi, C., Fagotto, F., \& Nasi, E.\ 1994, \aaps, 106, 275
\reference{} Bica, E.~L.~D., \& Schmitt, H.~R.\ 1995, \apjs, 101, 41
\reference{} Binney, J., Merrifield, M., 1998, Galactic astronomy. Princeton University Press, Princeton, NJ
\reference{} Binney, J., \& Tremaine, S., 1987, Galactic dynamics. Princeton University Press, Princeton, NJ
\reference{} Crowl, H.~H., Sarajedini, A., Piatti, A.~E., Geisler, D., Bica, E., Clari{\'a}, J.~J., \& Santos, J.~F.~C., Jr.\ 2001, \aj, 122, 220
\reference{} Da Costa, G.~S.\ 1991, IAU Symp.~148: The Magellanic Clouds, 148, 183
\reference{} Da Costa, G.~S., \& Hatzidimitriou, D.\ 1998, \aj, 115, 1934
\reference{} de Freitas Pacheco, J.~A., Barbuy, B., \& Idiart, T.\ 1998, \aap, 332, 19
\reference{} de Grijs, R., Gilmore, G.~F., Johnson, R.~A., \& Mackey, A.~D.\ 2002, \mnras, 331, 245
\reference{} Dolphin, A.~E.\ 2000, \pasp, 112, 1383
\reference{} Dolphin, A.~E., Walker, A.~R., Hodge, P.~W., Mateo, M., Olszewski, E.~W., Schommer, R.~A., \& Suntzeff, N.~B.\ 2001, \apj, 562, 303
\reference{} Elson, R.~A.~W., Fall, S.~M., \& Freeman, K.~C.\ 1987, \apj, 323, 54
\reference{} Girardi, L., et al.\ 2002, \aap, 391, 195
\reference{} Gouliermis, D., Keller, S.~C., Kontizas, M., Kontizas, E., \& Bellas-Velidis, I.\ 2004, \aap, 416, 137
\reference{} Gouliermis, D.~A., Dolphin, A.~E., Brandner, W., \& Henning, T.\ 2006, \apjs, 166, 549 (Paper I)
\reference{} Henize, K.~G.\ 1956, \apjs, 2, 315
\reference{} Kayser, A., Grebel, E.~K., Harbeck, D.~R., Cole, A.~A., Koch, A., Gallagher, J.~S., \& Da Costa, G.~S.\ 2006, ArXiv Astrophysics e-prints, arXiv:astro-ph/0607047
\reference{} Kerber, L.~O., \& Santiago, B.~X.\ 2006, \aap, 452, 155
\reference{} Kroupa, P.\ 2002, Science, 295, 82
\reference{} King, I.\ 1962, \aj, 67, 471
\reference{} Lightman, A.~P., \& Shapiro, S.~L. \ 1978, Reviews of Morden Physics, 50, 437
\reference{} McLaughlin, D.~E., \& van der Marel, R.~P.\ 2005, \apjs, 161, 304
\reference{} Mackey, A.~D., \& Gilmore, G.~F.\ 2003a, \mnras, 338, 85
\reference{} Mackey, A.~D., \& Gilmore, G.~F.\ 2003b, \mnras, 338, 120
\reference{} Mackey, A.~D., \& Gilmore, G.~F.\ 2003c, \mnras, 340, 175
\reference{} Mackey, A.~D., Payne, M.~J., \& Gilmore, G.~F.\ 2006, \mnras, 369, 921
\reference{} Meylan, G., \& Heggie, D.~C.\ 1997, \aapr, 8, 1
\reference{} Mighell, K.~J., Sarajedini, A., \& French, R.~S.\ 1998, \aj, 116, 2395
\reference{} Naz{\'e}, Y., Hartwell, J.~M., Stevens, I.~R., Manfroid, J., Marchenko, S., Corcoran, M.~F., Moffat, A.~F.~J., \& Skalkowski, G.\ 2003, \apj, 586, 983
\reference{} Naz{\'e}, Y., Manfroid, J., Stevens, I.~R., Corcoran, M.~F., \& Flores, A.\ 2004, \apj, 608, 208
\reference{} Pagel, B.~E.~J., \& Tautvai\v sien\'e, G.\ 1998, \mnras, 299, 535
\reference{} Peimbert, M., Peimbert, A., \& Ruiz, M.~T.\ 2000, \apj, 541, 688
\reference{} Piatti, A.~E., Santos, J.~F.~C., Clari{\'a}, J.~J., Bica, E., Sarajedini, A., \& Geisler, D.\ 2001, \mnras, 325, 792
\reference{} Piatti, A.~E., Sarajedini, A., Geisler, D., Seguel, J., \& Clark, D.\ 2005a, \mnras, 358, 1215
\reference{} Piatti, A.~E., Santos, J.~F.~C., Jr., Clari{\'a}, J.~J., \, E., Ahumada, A.~V., \& Parisi, M.~C.\ 2005b, \aap, 440, 111
\reference{} Pilyugin L. S., 1996, A\&A, 310, 751
\reference{} Rafelski, M., \& Zaritsky, D.\ 2005, \aj, 129, 2701
\reference{} Rich, R.~M., Shara, M., Fall, S.~M., \& Zurek, D.\ 2000, \aj, 119, 197
\reference{} Rubio, M., Contursi, A., Lequeux, J., Probst, R., Barb{\'a}, R., Boulanger, F., Cesarsky, D., \& Maoli, R.\ 2000, \aap, 359, 1139
\reference{} Russell S. C., \& Dopita M. A., 1992, ApJ, 384, 508
\reference{} Sabbi, E., et al.\ 2007, \aj, 133, 44
\reference{} Salpeter, E.~E.\ 1955, \apj, 121, 161
\reference{} Scalo, J.~M.\ 1986, Fundamentals of Cosmic Physics, 11, 1
\reference{} Spitzer, L., Jr.\ 1975, IAU Symp.~ 69: Dynamics of the Solar Systems, 69, 3
\reference{} Spitzer, L.\ 1987, Princeton, NJ, Princeton University Press, p. 191
\reference{} Tsujimoto T., Nomoto K., Yoshii Y., Hashimoto M., Yanagida S., Thielemann F.-K., 1995, MNRAS, 277, 945
\reference{} van den Bergh, S.\ 1991, \apj, 369, 1
\reference{} Vesperini, E., \& Heggie, D.~C.\ 1997, \mnras, 289, 898
\reference{} Westerlund, B.~E.\ 1997, The Magellanic Clouds. Cambridge University Press, Cambridge

\end{references}
\end{document}